\documentclass[letter, longauth]{aa} 
\bibpunct{(}{)}{;}{a}{}{,} 
\usepackage{graphicx}
\usepackage{txfonts}
\begin{document}

   \title{ALMA observations of anisotropic dust mass loss in the inner circumstellar environment of the red supergiant VY CMa}


   \author{E. O'Gorman
          \inst{1}\fnmsep\thanks{eamon.ogorman@chalmers.se},
          W. Vlemmings
          \inst{1},
          A. M. S. Richards
          \inst{2},
          A. Baudry
          \inst{3,4},
          E. De Beck          
          \inst{1},
          L. Decin
          \inst{5},
          G. M. Harper
          \inst{6},
		  E. M. Humphreys
          \inst{7},
          P. Kervella
          \inst{8,9,10},
          T. Khouri
          \inst{11},
          \and
          S. Muller
          \inst{1}
          }

   \institute{Department of Earth and Space Sciences, Chalmers University of Technology, Onsala Space Observatory, 439 92 Onsala, Sweden
          \and
              Jodrell Bank Centre for Astrophysics, School of Physics and Astronomy, University of Manchester, Manchester M13 9PL, UK
         \and
              Univ. Bordeaux, LAB, UMR 5804, F-33270 Floirac, France
         \and
	          CNRS, LAB, UMR 5804, F-33270 Floirac, France
         \and
              Instituut voor Sterrenkunde, Katholieke Universiteit Leuven, Celestijnenlaan 200D, 3001 Leuven, Belgium
         \and
              School of Physics, Trinity College Dublin, Dublin 2, Ireland
         \and
              ESO Karl-Schwarzschild-Str. 2, 85748 Garching, Germany
		 \and          
		      LESIA, Observatoire de Paris, CNRS, UPMC, Universit\'e Paris-Diderot, PSL, 5 place Jules Janssen, 92195 Meudon, France
         \and 
              UMI-FCA, CNRS/INSU, France (UMI 3386)
         \and
			  Dept. de Astronom\'ia, Universidad de Chile, Santiago, Chile
		 \and
		      Astronomical Institute Anton Pannekoek, University of Amsterdam, PO Box 94249, 1090 GE Amsterdam, The Netherlands
             }
 
  \abstract{The processes leading to dust formation and the subsequent role it plays in driving mass loss in cool evolved stars is an area of intense study. Here we present high resolution ALMA Science Verification data of the continuum emission around the highly evolved oxygen-rich red supergiant VY CMa. These data enable us to study the dust in its inner circumstellar environment at a spatial resolution of 129\,mas at 321\,GHz and 59\,mas at 658\,GHz, thus allowing us to trace dust on spatial scales down to 11\,R$_{\star}$ (71\,AU). Two prominent dust components are detected and resolved. The brightest dust component, C, is located 334\,mas (61\,R$_{\star}$) South East of the star and has a dust mass of at least $2.5\times 10^{-4}\,$M$_{\odot}$. It has a dust emissivity spectral index of $\beta =-0.1$ at its peak, implying that it is optically thick at these frequencies with a cool core of $T_{d}\lesssim 100$\,K. Interestingly, not a single molecule in the ALMA data has emission close to the peak of this massive dust clump. The other main dust component, VY, is located at the position of the star and contains a total dust mass of $4.0 \times 10^{-5}\,$M$_{\odot}$. It also contains a weaker dust feature extending over $60\,$R$_{\star}$ to the North with the total component having a typical dust emissivity spectral index of $\beta =0.7$. We find that at least $17\%$ of the dust mass around VY CMa is located in clumps ejected within a more quiescent roughly spherical stellar wind, with a quiescent dust mass loss rate of $5 \times 10^{-6}$\,M$_{\odot}\,$yr$^{-1}$. The anisotropic morphology of the dust indicates a continuous, directed mass loss over a few decades, suggesting that this mass loss cannot be driven by large convection cells alone.}
   \titlerunning{ALMA Observations of the Dust Around VY CMa}
   \authorrunning{E. O'Gorman et al.}
   \maketitle

\section{Introduction}
Cool evolved stars are the main producers of dust in galaxies and are important drivers of the chemical evolution of matter in the Universe. Despite the importance of dust in a broad range of astrophysical phenomena, the conditions that lead to its formation in the outflows of evolved stars and its subsequent role in driving mass loss remain largely unknown. Oxygen-rich red supergiants (RSGs) are sources of inorganic dust (silicate and alumina) that is formed and plays a role in launching mass loss in the inner circumstellar environment of these stars. However, the dust properties in this region such as density, composition, and morphology have remained largely unknown because the spatial resolution is not adequate. 

A prime target for studying the properties of dust around evolved stars is the enigmatic oxygen-rich RSG VY Canis Majoris (VY CMa). VY CMa is a mid-M spectral type  \citep[M5e\,Ia;][]{humphreys_1974} RSG with an extremely high mass-loss rate \citep[$\dot{M} \sim 3 \times 10^{-4}\,M_{\odot}\,\rm{yr}^{-1}$;][]{danchi_1994} and can undergo periods of even more intense mass loss \citep{humphreys_2007}. This mass loss is about two orders of magnitude greater than the well known early-M spectral type RSGs, Betelgeuse and Antares. Consequently, VY CMa has a highly dense and dusty circumstellar envelope that obscures it and produces a reflection nebula at optical wavelengths \citep{humphreys_2007}. Thanks to its relative proximity \citep[$d=1.2^{+0.13}_{-0.10}\,$kpc;][]{zhang_2012} and high intrinsic luminosity ($L= 3 \times 10^5 \,L_{\odot}$ using the photometry of Smith et al., 2001), the dust emission from the circumstellar envelope around VY CMa has been studied from optical to centimeter wavelengths \citep[e.g., ][]{lipscy_2005,muller_2007}. The dusty envelope consists of a diffuse and extended region with loops and arcs expanding over several arc seconds through a more uniform medium. At the sub-arc-second level, the envelope consists of a dense and dusty central core, but little is known about it. It is within this region that the dust condenses and radiation pressure on dust helps drive mass-loss. In this Letter we report the results of a sub-100\,mas study of this region using ALMA Science Verification data.

\section{Observations and results} 
VY CMa was observed at 321~GHz, 325~GHz, and 658~GHz as part of the ALMA Science Verification process on 2013 16-19 August, using 16 to 20 antennas of the main array with projected baselines ranging from 14\,m to 2.7\,km. Details of the observations and data processing are provided in \cite{richards_2014}, hereafter R+14. For the purpose of the continuum analysis, we used a total of $1.74$~GHz line free continuum channels around 321~GHz and $0.4$~GHz centered on 658~GHz. As the data around 325~GHz provided similar results to the 321~GHz data, but were affected by the 325.15 GHz atmospheric water line resulting in increased noise, these data were not included in our subsequent analysis. The synthesized beam size, rms noise level, maximum recoverable scale (MRS), and total flux density of the two observing bands are given in Table~\ref{tab1}. The rms noise is not uniform across the field, most likely because of low surface brightness and resolved-out emission (see below). In Table~\ref{tab1} we report the rms in the region of maximum noise level, while in R+14, the rms corresponds to the lower rms areas. 

Uncertainties in the phase transfer of the data resulted in $\sim 110\,$mas offsets between the main continuum peaks in the  321 and 658\,GHz images. To find the optimal (RA, Dec)-offset correction, we shifted the 658\,GHz image relative the 321\,GHz image and computed a 2-D cross-correlation function for a range of shifts. The maximum correlation between the two images gave a (-80\,mas, -30\,mas)-offset correction for the 658\,GHz image, similar to the correction found by R+14 when aligning the peak position of the second brightest continuum component. The 321\,GHz image and the position corrected 658\,GHz image are plotted in white and black contours in Figure \ref{fig1}, respectively. Continuous emission was detected above the $3\sigma _{\mathrm{rms}}$ level on scales as large as $1.4\arcsec$ at 321\,GHz and $0.8\arcsec$ at 658\,GHz. 

There are two prominent components and a weaker extended N-NE feature that are reproduced at both frequencies, as seen in Figure \ref{fig1}. The position around the peak of the secondary component has recently been shown to coincide with the center of expansion of many maser emission lines and has been deduced to be the location of the star (R+14). This component (VY) connects to the weaker emission extending for $\sim 800$\,mas at a position angle (PA, measured east of north) of $\sim 20^\circ$, which may contain some further substructure. Interestingly, the brightest component in both images (C) is not at the location of the star itself but is 334\,mas SE of the star in the plane of the sky. Using a distance of 1.2\,kpc to VY CMa, this angular distance corresponds to 400\,AU, or 61\,R$_{\star}$ if a linear radius of 1420\,R$_{\odot}$ (at $2~\mu$m) is adopted (Wittkowski, 2012). R+14 have found that both blue- and redshifted 325\,GHz and 658\,GHz masers, which are emitted out to $\sim240$\,AU, straddle this main continuum component, so the projected distance between this component and the star is likely to be close to the actual distance.

  \begin{table}
         \caption[]{ALMA continuum observations of VY CMa.}
      \vspace{-8mm}
         \label{tab1}
     $$ 
         \begin{array}{ccccc}
            \hline
            \noalign{\smallskip}
            \nu & $Synthesized Beam$ & $rms noise$ & $MRS$ & $Total $ S_{\nu} \\
              $(GHz)$ & $($^{\prime\prime} \times ^{\prime\prime}$, PA)$ & $(mJy~beam$^{-1}$)$ & $(\arcsec)$ &$(mJy)$\\
            \noalign{\smallskip}
            \hline
            \noalign{\smallskip}
            321   & 0.229\times 0.129, 28^{\circ} & 0.6 & 8.3 &587\\
	    658   & 0.110\times 0.059, 30^{\circ} & 6 & 4.0 &3017\\
            \noalign{\smallskip}
            \hline
         \end{array}
     $$ 
      \vspace{-5mm}
   \end{table}

We fit 2-D elliptical Gaussian components to the two main continuum features and found that both the C and the VY components were resolved at both frequencies. The sizes and flux densities of these components were estimated from the deconvolved fits and are listed in Table \ref{tab2}. The main C component is highly elongated at both frequencies in the SE direction, with a major-to-minor axis ratio, $\theta _{\mathrm{maj}}/\theta _{\mathrm{min}} = 1.6\pm0.2$ at 321\,GHz and $\theta _{\mathrm{maj}}/\theta _{\mathrm{min}} = 2.1\pm0.1$ at 658\,GHz, with the axis of elongation pointing in the direction toward VY. We find that the size of component C is consistent at both frequencies within their errors. Comparing Tables \ref{tab1} and \ref{tab2}, we can see that 50$\%$ and 60$\%$ of the total continuum flux density emanates from the main C component at 321\,GHz and 658\,GHz, respectively. The other main continuum component VY, is also resolved at both frequencies. It is elongated toward the North with  $\theta _{\mathrm{maj}}/\theta _{\mathrm{min}} = 1.5\pm0.4$ at 321~GHz and $\theta _{\mathrm{maj}}/\theta _{\mathrm{min}} = 2.1\pm0.5$ at 658 GHz, with emission extending to 27\,R$_{\star}$ (180\,AU). It is centered on the location of the stellar photosphere, so it includes emission from the star itself at these sub-millimeter frequencies. The VY component contributes about 26$\%$ and 17$\%$ of the total ALMA continuum flux density at 321\,GHz and 658\,GHz, respectively. The significant difference in size at the two frequencies cannot be attributed to the MRS on these small angular scales, but is probably due to a combination of different sensitivities to low surface brightness emission and the fact that the higher frequencies probe emission from hotter regions of VY.

A spectral index map of the continuum emission was created after convolving the 658\,GHz data with the synthesized beam from the 321\,GHz image, and regridding the convolved image to match the pixel scale of the 321\,GHz image. The resulting spectral index map is represented by the color image in Figure \ref{fig2}, while the contour levels represent the convolved 658\,GHz image. All data below the $6\sigma _{\mathrm{rms}}$ noise level in the convolved 658\,GHz image have been omitted from Figure \ref{fig2}. The middle region of the main continuum component C has a spectral index of $\alpha \sim 1.9$ (where $S_{\nu} \propto \nu^{\alpha}$). The spectral index value around the second continuum, or VY, and most of the extended emission in the N-NE direction has a spectral index of $\alpha \sim 2.7$. The statistical error on the spectral index due to random noise errors is less than 0.2 in the $6\sigma _{\mathrm{rms}}$ region and decreases to 0.05 in the peak region. The S/N at 658\,GHz is less than that at 321\,GHz, so this error is dominated by the 658\,GHz data. There is also a systematic absolute error of $\sim 0.22$ on the entire spectral index map based on the $\pm 15\%$ uncertainty of the absolute flux calibration (R+14). We note that for VY, the difference in spectral index between the pixel flux densities in the map and what can be derived from the integrated flux densities is due to the smaller size of VY at 658~GHz.

\section{Discussion and conclusions}
\subsection{Comparison with \textit{HST}}
\begin{figure}
   \centering
   \includegraphics[trim = 0mm 15mm 10mm 0mm, clip,scale=0.45]{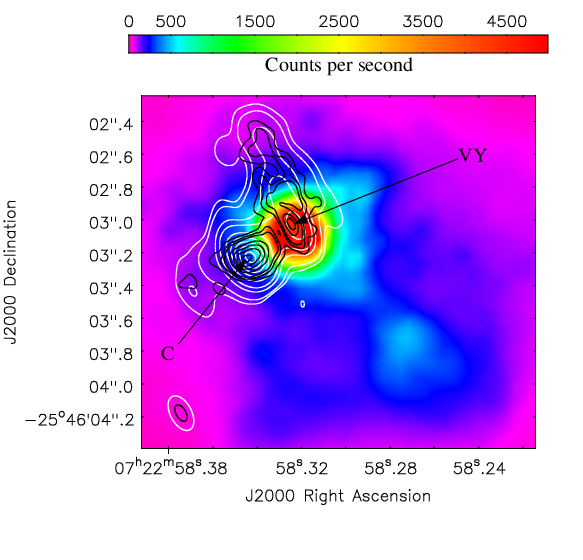}
      \caption{Color image is an \textit{HST}/WFPC2 exposure of VY CMa using the F1042M filter (Smith et al., 2001) corrected for proper motion. The data has been truncated at 5000 counts per second and scaled with a power cycle of -1.5 to highlight the extended emission in the SW direction. The contours represent the ALMA data at 321\,GHz (white) and 658\,GHz (black) and the corresponding synthesized beams are located in the bottom left of the image. The 321\,GHz contour levels are set at $[5,10,30,50,100,....,300]\times \sigma_{\mathrm{rms}}$, while the 658\,GHz contour levels are set at $[3,6,9,20,30,40,50]\times \sigma_{\mathrm{rms}}$.
              }
         \label{fig1}
   \end{figure}

The color image in Figure \ref{fig1} is an \textit{HST}/WFPC2 exposure of VY CMa taken with the F1042M filter on 1999 March 22 \citep{smith_2001}. The data is adjusted for proper motion using the values of \cite{zhang_2012}, and it can be seen that the peak emission fits the second main continuum component VY very well. The typical positional uncertainty rms in this \textit{HST} image is 48\,mas in RA and 146\,mas in Dec, while the ALMA positional error is 35\,mas, ruling out a correspondence between the C component in the ALMA images and the peak of the \textit{HST} image. We can conclude that most of the escaping scattered light at optical wavelengths is emitted along the line of sight to the star itself. There is no evidence of the main C component in the \textit{HST} image, probably because the \textit{HST} emission does not trace dense dusty emission, but rather more tenuous regions in the circumstellar environment. Scattered light polarimetry with the \textit{HST} does show a pronounced lower polarization toward the SE of the star \citep{jones_2007}, potentially indicating enhanced obscuration along the direction between the star and component C. 

The extended SW emission feature in the \textit{HST} image, which may be a density cavity in the circumstellar environment formed by a previous episodic mass-loss event, has no detectable signature in our ALMA continuum images. This extended feature ends in what is named the SW clump in \cite{smith_2001} and \cite{humphreys_2007}. This clump is thought to be related to an ejection event from the star and has been estimated to have a dust-mass lower limit of $5\times10^{-5}~M_{\odot}$ and a temperature between 80~K and 210~K \citep{shenoy_2013}. However, we see no sign of this clump in the ALMA data. With our $3\sigma_{\rm rms}$ limits, we rule out a mass $>7\times10^{-6}~M_{\odot}$ (for $T=80$~K) or $>3\times10^{-6}~M_{\odot}$ (for $T=210$~K). For the ALMA observations to be consistent with the lower limit estimates of \cite{shenoy_2013}, the dust temperature would have to be $<20$~K. Alternatively, a different dust composition with different dust scattering and emissivity indices needs to be invoked. The exact nature of the SW clump is thus not clear. We also find no other compact submillimeter emission in the larger $\sim8\arcsec$ region of dust seen with the \textit{HST}.

%
 
   \begin{figure}
   \centering
   \includegraphics[trim = 25mm 120mm 35mm 30mm, clip,scale=0.54, angle=0]{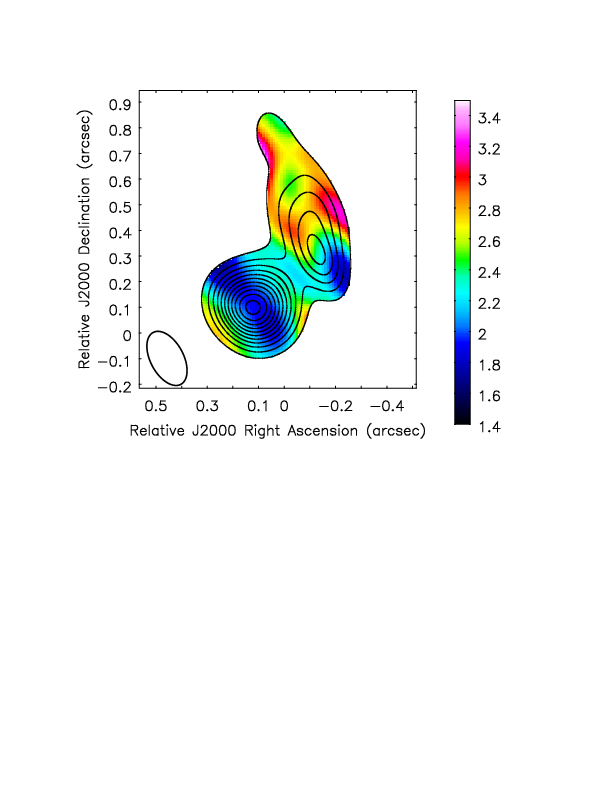}
      \caption{Color image shows the spectral index map derived from the 321\,GHz and 658\,GHz ALMA continuum maps. The strongest continuum component has a spectral index $\alpha = 1.9\pm 0.05$ at the peak, and the weaker elongated emission has a typical spectral index $\alpha = 2.7\pm 0.1$. The error values here do not include the systematic error based on the $\pm 15\%$ uncertainty of the absolute flux calibration, which is in the same direction for both regions of the image. The contour map represents the 658\,GHz map, which was convolved with the 321\,GHz restoring beam to create the spectral index map. Contour levels are set at $[6,9,...,39]\times \sigma_{\mathrm{rms}}$.}
         \label{fig2}
   \end{figure}
%

\subsection{Properties of the dust}
\label{sec3.2}
Dust is the main source of the observed thermal submillimeter emission around VY CMa. If the dust is optically thin at these frequencies, and we assume the Rayleigh-Jeans approximation, then the observed flux density minus the stellar flux density contribution $S_{\star}$ is proportional to the mass of the emitting dust $M_d$, such that
\begin{equation}
S_{\nu} - S_{\star}=\frac{3M_{d}Q_{\nu}T_{d}k\nu^2}{2a_{g}\rho _{g}c^2d^2}
\label{eq1}
\end{equation}
where $Q _{\nu}$ is the grain emissivity, $a_{g}$ and $\rho _{g}$ are the radius and mass density of the dust grains, respectively, $d$ is the distance to the star, and $T_d$ is the dust temperature \citep{knapp_1993}. Assuming that the grain emissivity has a power law dependence on frequency, $Q (\nu) \propto \nu ^{\beta}$, then the flux density is $S_{\nu} \propto \nu ^{\alpha}$, where $\alpha = 2 + \beta$. Here, $\beta$ is the dust emissivity spectral index and has typical values of $\beta = 1.8 \pm 0.2$ for both diffuse and dense clouds in the interstellar medium \citep{draine_2006}. For evolved stars, the dust emissivity index is generally lower \citep{knapp_1993}, and for VY CMa previous observations have found values between $\beta \sim 0-1.0$ (\citealt{shinnaga_2004}, \citealt{kaminski_2013}), while its circumstellar environment has been modeled using values of $\beta = 0.9$ \citep{knapp_1993}.

Our spectral index map in Figure \ref{fig2} shows that the two spatially resolved continuum components in the inner circumstellar envelope have different spectral indices. The second brightest continuum component, VY, has a typical dust emissivity index of $\beta = 0.7\pm 0.1$ and is consistent with the previous single-dish observations outlined in Appendix \ref{app1}. However, much of the main continuum component, C, has a dust emissivity index of $\beta = -0.1$ around the location of the peak emission, meaning that the submillimeter emission from this component is, or is close to, becoming optically thick. Following \cite{herman_1986}, optically thin dust at a radius $r_d$ is heated by the incident stellar radiation field to a dust temperature
\begin{equation}
T_{d} = \left(\frac{L_{\star}T_{\star}^{\beta}}{16\pi\sigma r_{d}^2}\right)^{1/(4+\beta)}
\end{equation}
where $L_{\star}$ and $T_{\star}$ are the stellar luminosity and temperature, respectively. Assuming $T_{\star} = 3490\,$K \citep{wittkowski_2012}, the main continuum component, C, would have an isothermal temperature of 450\,K at 400\,AU from the star, while the VY component would have a value of 970\,K at a mean dust radius of 67\,AU. We stress that in deriving the temperature for the C component, we have assumed that the dust is optically thin even though the spectral index map indicates otherwise. We thus use 450\,K as an upper limit to the temperature of C. These rough estimates for the dust temperature allow us to empirically calculate the optical depth at both observing frequencies, the values of which are given in Table \ref{tab2}. The VY component is optically thin at both frequencies with $\tau _{321-658\,\mathrm{GHz}}=0.02-0.06$, while the main continuum component C, has an optical depth of $\tau _{321-658\,\mathrm{GHz}} \sim 0.2$, and would become optically thick at $110-120\,$K at both frequencies. Therefore, a cooler dust component in the center of C would explain the spectral index having a lower value than expected for optically thin dust. Alternatively, larger dust grains could also result in low $\beta$ values although our observing frequencies would then suggest that the dust grains would need to be much larger than $100\,\mu$m. Interestingly, we find that the thermal molecular emission lines from the ALMA data avoids the C component, as is consistent with the molecular emission offsets found in SMA observations \citep{muller_2007, kaminski_2013}. This could potentially support the low temperature and high optical depth suggestion because of low excitation and/or depletion onto dust grains.

Equation \ref{eq1} can be re-arranged to find the total mass of dust in each of the two main continuum components. The main component C may not be fully optically thin at both frequencies and so its dust mass estimates will be lower limits. Following \cite{knapp_1993} we assume typical oxygen-rich star values for the radius and mass density of the dust grains, $a_{g} = 0.2\,\mu$m and $\rho _{g}=3.5\,$g\,cm$^{-3}$. We slightly alter their grain emissivity function to include our derived dust emissivity spectral index so that $Q_{\nu} = 5.65\times 10^{-4}(\nu /274.6\,\mathrm{GHz})^{0.7}$. We also assume that the main C component has the same dust composition as VY, so the dust emissivity spectral index of C would also be $\beta = 0.7$ if fully optically thin. We then derive dust masses of $2.5\times 10^{-4}\,M_{\odot}$ at 321\,GHz and $1.6\times 10^{-4}\,M_{\odot}$ at 658\,GHz for the main dust component C, having assumed no photospheric contribution. The lower mass value at 658\,GHz is probably due to a combination of optical depth effects and an inaccurate grain emissivity law. We derive dust masses of $4.0\times 10^{-5}\,M_{\odot}$ at 321 GHz and $1.8\times 10^{-5}\,M_{\odot}$ at 658 GHz for the VY component having subtracting the stellar contribution, which is calculated in Appendix \ref{app2}. The discrepancy in the dust mass values here could be due to an inaccurate grain emissivity law or differences in sensitivities to low surface brightness emission.
 
\subsection{Dust mass-loss history and mechanisms}
Considering that we find that part of the dust emission around VY~CMa arises in compact clumps, we can obtain a new estimate of the average dust mass-loss rate by determining the dust mass that is resolved out in the ALMA observations. We assume the dust to be extended over the $\sim8\arcsec$ size of the \textit{HST} image and use an average dust velocity $V_{\rm{d}}$ that depends on grain size and that for VY CMa is $\sim 60$~km\,s$^{-1}$ for grains of $\sim 0.25\,\mu$m \citep{decin_2006}. We assume an average temperature of 200\,K throughout the shell. The observed VY component arises from the dust lost in the last $\sim 20(60~$km\,s$^{-1}/V_{\rm{d}})$~yr and the resolved-out emission ($\sim 1.1$~Jy at 321~GHz and $\sim 7.8$~Jy at 658~GHz as shown in Figure \ref{fig3}) corresponds to the dust lost over $\sim 360 (60~$km\,s$^{-1}/V_{\rm{d}})$~yr before. For $\beta=0.7$ using Eq. \ref{eq1}, we find a dust mass of $\sim1.8\times10^{-3}\,M_{\odot}$ which is consistent at both frequencies, and a dust mass-loss rate of $5\times10^{-6}(V_{\rm{d}}/60~$km\,s$^{-1})\,$M$_{\odot}\,$yr$^{-1}$, which is four times higher than previously estimated by \cite{sopka_1985} and slightly higher than \cite{knapp_1993}, taking the different adopted velocities and distances into account. Then, taking all compact emission from the ALMA observations, with the exception of what is around the star, to be in optically thin clumps at $\sim450$~K, those clumps contain $\sim3\times10^{-4}\,M_{\odot}$ of dust. This is a lower limit, because a significant component of cold dust can potentially be present in C and in the SW and other \textit{HST} features not detected by ALMA. We can thus conclude that $>17\%$ of the dust mass around VY~CMa is located in clumps ejected within a more quiescent, roughly spherical stellar wind. This picture of a quiescent mass loss disrupted with periods of more localized intense ejections of mass is consistent with the findings of \cite{humphreys_2007} for VY CMa and \cite{ohnaka_2014} for Antares (M1~Iab).
 
It has been hypothesized that large convection cells initiate mass loss in RSGs by lifting material out to distances beyond the photosphere where dust can form and drive mass loss \citep[e.g.,][]{lim_1998}. These convection cells are predicted to have time scales of $\sim 150\,$days \citep{schwarzschild_1975} and are expected to eject material at random PAs from the star, generating supersonic motions and shocks in the process \citep{lion_2013}. Our observations have revealed a massive dust clump centered at 60\,R$_{\star}$ and a weaker continuous dust feature extending 90\,R$_{\star}$ to the North. These features imply a continuous directed dust mass-loss from the star over the past $30-50\,$yr. This contrasts with what would be expected from mass loss that is only driven by convection, which on these spatial scales would show many dust clumps at random PAs. A mass-loss mechanism that is localized but much more stable over time (i.e., over a few decades) is required to explain our observations. This would suggest a magnetic origin. Indeed a surface magnetic field strength of $10^3$\,G has been extrapolated from circumstellar maser measurements for VY CMa \citep{vlemmings_2002}. Cool spots on the photosphere due to localized long-lived MHD disturbances could then enhance local dust formation, and hence drive mass loss in the localized directions that are evident in our ALMA data. Such a mass-loss mechanism would still appear as random mass ejections on larger spatial scales and could be the source of the arcs and clumps seen in VY CMa's outer circumstellar environment by \cite{smith_2001} and \cite{humphreys_2007}.

\begin{acknowledgements}
This paper makes use of the following ALMA data: ADS/JAO.ALMA\#2011.0.00011.SV. ALMA is a partnership of ESO (representing its member states), NSF (USA) and NINS (Japan), together with NRC (Canada) and NSC and ASIAA (Taiwan), in cooperation with the Republic of Chile. The Joint ALMA Observatory is operated by ESO, AUI/NRAO and NAOJ. EOG and WV acknowledge support from Marie Curie Career Integration Grant 321691 and ERC consolidator grant 614264
\end{acknowledgements}

\bibliographystyle{aa}
\bibliography{references}

\begin{thebibliography}{25}
\expandafter\ifx\csname natexlab\endcsname\relax\def\natexlab#1{#1}\fi

\bibitem[{{Danchi} {et~al.}(1994){Danchi}, {Bester}, {Degiacomi}, {Greenhill},
  \& {Townes}}]{danchi_1994}
{Danchi}, W.~C., {Bester}, M., {Degiacomi}, C.~G., {Greenhill}, L.~J., \&
  {Townes}, C.~H. 1994, \aj, 107, 1469

\bibitem[{{Decin} {et~al.}(2006){Decin}, {Hony}, {de Koter}, {Justtanont},
  {Tielens}, \& {Waters}}]{decin_2006}
{Decin}, L., {Hony}, S., {de Koter}, A., {et~al.} 2006, \aap, 456, 549

\bibitem[{{Draine}(2006)}]{draine_2006}
{Draine}, B.~T. 2006, \apj, 636, 1114

\bibitem[{{Harper} {et~al.}(2001){Harper}, {Brown}, \& {Lim}}]{harper_2001}
{Harper}, G.~M., {Brown}, A., \& {Lim}, J. 2001, \apj, 551, 1073

\bibitem[{{Herman} {et~al.}(1986){Herman}, {Burger}, \&
  {Penninx}}]{herman_1986}
{Herman}, J., {Burger}, J.~H., \& {Penninx}, W.~H. 1986, \aap, 167, 247

\bibitem[{{Humphreys}(1974)}]{humphreys_1974}
{Humphreys}, R.~M. 1974, \apj, 188, 75

\bibitem[{{Humphreys} {et~al.}(2007){Humphreys}, {Helton}, \&
  {Jones}}]{humphreys_2007}
{Humphreys}, R.~M., {Helton}, L.~A., \& {Jones}, T.~J. 2007, \aj, 133, 2716

\bibitem[{{Jones} {et~al.}(2007){Jones}, {Humphreys}, {Helton}, {Gui}, \&
  {Huang}}]{jones_2007}
{Jones}, T.~J., {Humphreys}, R.~M., {Helton}, L.~A., {Gui}, C., \& {Huang}, X.
  2007, \aj, 133, 2730

\bibitem[{{Kami{\'n}ski} {et~al.}(2013){Kami{\'n}ski}, {Gottlieb}, {Young},
  {Menten}, \& {Patel}}]{kaminski_2013}
{Kami{\'n}ski}, T., {Gottlieb}, C.~A., {Young}, K.~H., {Menten}, K.~M., \&
  {Patel}, N.~A. 2013, \apjs, 209, 38

\bibitem[{{Knapp} {et~al.}(1993){Knapp}, {Sandell}, \& {Robson}}]{knapp_1993}
{Knapp}, G.~R., {Sandell}, G., \& {Robson}, E.~I. 1993, \apjs, 88, 173

\bibitem[{{Ladjal} {et~al.}(2010){Ladjal}, {Justtanont}, {Groenewegen},
  {Blommaert}, {Waelkens}, \& {Barlow}}]{ladjal_2010}
{Ladjal}, D., {Justtanont}, K., {Groenewegen}, M.~A.~T., {et~al.} 2010, \aap,
  513, A53

\bibitem[{{Lim} {et~al.}(1998){Lim}, {Carilli}, {White}, {Beasley}, \&
  {Marson}}]{lim_1998}
{Lim}, J., {Carilli}, C.~L., {White}, S.~M., {Beasley}, A.~J., \& {Marson},
  R.~G. 1998, \nat, 392, 575

\bibitem[{{Lion} {et~al.}(2013){Lion}, {Van Eck}, {Chiavassa}, {Plez}, \&
  {Jorissen}}]{lion_2013}
{Lion}, S., {Van Eck}, S., {Chiavassa}, A., {Plez}, B., \& {Jorissen}, A. 2013,
  in EAS Publications Series, Vol.~60, EAS Publications Series, ed.
  P.~{Kervella}, T.~{Le Bertre}, \& G.~{Perrin}, 85--92

\bibitem[{{Lipscy} {et~al.}(2005){Lipscy}, {Jura}, \& {Reid}}]{lipscy_2005}
{Lipscy}, S.~J., {Jura}, M., \& {Reid}, M.~J. 2005, \apj, 626, 439

\bibitem[{{Muller} {et~al.}(2007){Muller}, {Dinh-V-Trung}, {Lim}, {Hirano},
  {Muthu}, \& {Kwok}}]{muller_2007}
{Muller}, S., {Dinh-V-Trung}, {Lim}, J., {et~al.} 2007, \apj, 656, 1109

\bibitem[{{Ohnaka}(2014)}]{ohnaka_2014}
{Ohnaka}, K. 2014, \aap, 568, A17

\bibitem[{{Richards} {et~al.}(2014){Richards}, {Impellizzeri}, {Humphreys},
  {Vlahakis}, {Vlemmings}, {Baudry}, {De Beck}, {Decin}, {Etoka}, {Gray},
  {Harper}, {Hunter}, {Kervella}, {Kerschbaum}, {McDonald}, {Melnick},
  {Muller}, {Neufeld}, {O'Gorman}, {Parfenov}, {Peck}, {Shinnaga}, {Sobolev},
  {Testi}, {Uscanga}, {Wootten}, {Yates}, \& {Zijlstra}}]{richards_2014}
{Richards}, A.~M.~S., {Impellizzeri}, C.~M.~V., {Humphreys}, E.~M., {et~al.}
  2014, ArXiv e-prints

\bibitem[{{Schwarzschild}(1975)}]{schwarzschild_1975}
{Schwarzschild}, M. 1975, \apj, 195, 137

\bibitem[{{Shenoy} {et~al.}(2013){Shenoy}, {Jones}, {Humphreys}, {Marengo},
  {Leisenring}, {Nelson}, {Wilson}, {Skrutskie}, {Hinz}, {Hoffmann}, {Bailey},
  {Skemer}, {Rodigas}, \& {Vaitheeswaran}}]{shenoy_2013}
{Shenoy}, D.~P., {Jones}, T.~J., {Humphreys}, R.~M., {et~al.} 2013, \aj, 146,
  90

\bibitem[{{Shinnaga} {et~al.}(2004){Shinnaga}, {Moran}, {Young}, \&
  {Ho}}]{shinnaga_2004}
{Shinnaga}, H., {Moran}, J.~M., {Young}, K.~H., \& {Ho}, P.~T.~P. 2004, \apjl,
  616, L47

\bibitem[{{Smith} {et~al.}(2001){Smith}, {Humphreys}, {Davidson}, {Gehrz},
  {Schuster}, \& {Krautter}}]{smith_2001}
{Smith}, N., {Humphreys}, R.~M., {Davidson}, K., {et~al.} 2001, \aj, 121, 1111

\bibitem[{{Sopka} {et~al.}(1985){Sopka}, {Hildebrand}, {Jaffe}, {Gatley},
  {Roellig}, {Werner}, {Jura}, \& {Zuckerman}}]{sopka_1985}
{Sopka}, R.~J., {Hildebrand}, R., {Jaffe}, D.~T., {et~al.} 1985, \apj, 294, 242

\bibitem[{{Vlemmings} {et~al.}(2002){Vlemmings}, {Diamond}, \& {van
  Langevelde}}]{vlemmings_2002}
{Vlemmings}, W.~H.~T., {Diamond}, P.~J., \& {van Langevelde}, H.~J. 2002, \aap,
  394, 589

\bibitem[{{Wittkowski} {et~al.}(2012){Wittkowski}, {Hauschildt},
  {Arroyo-Torres}, \& {Marcaide}}]{wittkowski_2012}
{Wittkowski}, M., {Hauschildt}, P.~H., {Arroyo-Torres}, B., \& {Marcaide},
  J.~M. 2012, \aap, 540, L12

\bibitem[{{Zhang} {et~al.}(2012){Zhang}, {Reid}, {Menten}, \&
  {Zheng}}]{zhang_2012}
{Zhang}, B., {Reid}, M.~J., {Menten}, K.~M., \& {Zheng}, X.~W. 2012, \apj, 744,
  23

\end{thebibliography}

\begin{appendix}
\section{Properties of the two main continuum components}
\label{app0}
In Table~\ref{tab2} we list the properties of the two main continuum components, C and VY. 

      \begin{table*}
      \centering
         \caption[]{Properties of the two main continuum components, C and VY.}
      \vspace{-6mm}
         \label{tab2}
     $$ 
         \begin{array}{cccccccccccc}
            \hline
            \noalign{\smallskip}
            $ID$_{\nu}     &  \theta_{\mathrm{maj}} & \theta_{\mathrm{maj}}&\theta_{\mathrm{min}} & \theta_{\mathrm{min}}&$PA$ &  S_{\nu}& \alpha & \beta &T_\mathrm{d} & \tau _{\nu} & M_\mathrm{d}\\
                       &  $(mas)$ & $(R$_{\star}$)$& $(mas)$ & $(R$_{\star}$)$ &$($^{\circ}$)$ &$(mJy)$ & & &$(K)$ &  & $($M_{\odot}$)$\\
            \noalign{\smallskip}
            \hline
            \noalign{\smallskip}
            $C$_{321} & 182\pm 13 &33& 117\pm 16 &21&112\pm 10 &348\pm 16 &1.9 & -0.1&<450& 0.19&2.5\times 10^{-4}\\
	        $C$_{658} &  206\pm4  &37& 99\pm3 & 18&115\pm 1&1532\pm 55&  1.9& -0.1&<450& 0.17&1.6\times 10^{-4}\\
            $VY$_{321} &  219\pm19  &40& 144\pm 33 &26&4\pm 17 &155\pm 14 &2.7 & 0.7&970& 0.04 &4.0\times 10^{-5}\\
	        $VY$_{658} & 150\pm 8 &27& 73\pm 17 & 13&22\pm 5 &501\pm53 & 2.7& 0.7&970 & 0.02&1.8\times 10^{-5}\\
            \noalign{\smallskip}
            \hline
         \end{array}
     $$ 
      \vspace{-5mm}
     \tablefoot{$\theta_{\mathrm{maj}}$ and $\theta_{\mathrm{min}}$: Major and minor axes of the deconvolved components of the Gaussian fits to the continuum components; PA: position angle; $S_{\nu}$: integrated flux density from the deconvolved fits; $\alpha$: spectral index; $\beta$: dust emissivity spectral index; $T_{d}$: dust temperature assuming isothermal and optically thin dust, which may not be the case for the C component; $\tau _{\nu}$: optical depth; $M_{d}$: dust mass.}
   \end{table*}

\section{Comparison with other (sub-)mm observations}
\label{app1}
\begin{figure}
   \centering
   \includegraphics[trim = 0mm 0mm 0mm 10mm, clip,scale=0.4, angle=90]{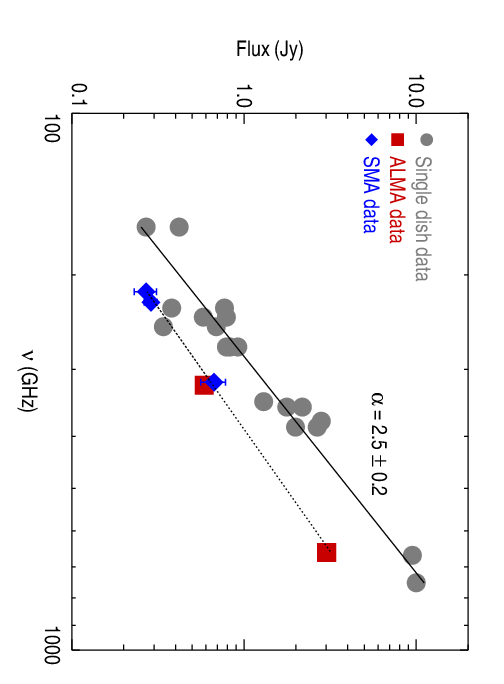}
      \caption{Compilation of the single dish bolometer and interferometric mm/sub-mm continuum observations of VY~CMa. The gray filled circles are single-dish flux density measurements, while the blue filled diamonds and the red filled squares are interferometric measurements. The solid line indicates the spectral index fit to the single-dish observations, with $\alpha=2.5\pm0.2$, while the dotted line is a linear fit to the interferometric observations.}
         \label{fig3}
   \end{figure}

There have been many previous (sub-)millimeter continuum flux density measurements of VY~CMa. In Fig.~\ref{fig3}, we present both the single-dish observations, using the compilations of \cite{knapp_1993} and \cite{ladjal_2010}, and the interferometric measurements from this paper and from SMA observations \citep{shinnaga_2004, muller_2007, kaminski_2013}. It is immediately apparent that the interferometric observations systematically underestimate the total flux density and can be attributed to flux density that is resolved out by missing short baselines, as well as by having low surface brightness. The MRS at higher frequencies is smaller so more flux density is resolved out, resulting in a larger discrepancy between the single-dish and interferometric measurements. The spectral index determined from the single-dish observations  is $\alpha=2.5\pm0.2$ and is consistent with our findings in Section \ref{sec3.2} for the optically thin dust. We note that the single-dish data points in Figure \ref{fig3} will also contain flux density from molecular emission lines. \cite{kaminski_2013} find that this accounts for about 25\% of their measured flux density between 279 and 355\,GHz. This level will vary at the different frequencies and will mainly result in an added upward scatter but should not affect the overall spectral index. Assuming the true spectral index of the interferometric observations to be the same as that of the single-dish observations and assuming there are no significant changes in dust properties on the different scales, we can conclude that the ALMA observations lose $\sim65\%$ of the emission at 321~GHz and $\sim72\%$ of the emission at 658~GHz. Considering the MRS and sensitivity of the ALMA observations, most of the submillimeter dust continuum flux density is thus located in a smooth low surface brightness distribution stretching beyond $\sim4\arcsec$, which is consistent with the dust distribution seen over $\sim8\arcsec$ with the \textit{HST} \citep[e.g.,][]{smith_2001}.

\label{comparison}
   
\section{Stellar flux density contribution}
\label{app2}
The stellar flux density contribution $S_{\star}$, needs to be considered when calculating the dust mass from the VY component at these ALMA frequencies. Previous studies have estimated this contribution by calculating the flux density of an optically thick blackbody of radius $R_{\star}$ and temperature $T_{\mathrm{eff}}$. Estimating the contribution from this method yields $S_{\star} = 26.5\,$mJy at 321\,GHz and $S_{\star} = 111\,$mJy at 658\,GHz, where we have assumed a stellar radius of $R_{\star}= 1420\,R_{\odot}$ (6.84\,AU or 5.7\,mas) and an effective temperature $T_{\mathrm{eff}}= 3490\,$K \citep{wittkowski_2012}. However, RSGs have weakly ionized extended atmospheres, which will become opaque at (sub-)mm frequencies. To estimate this contribution at these ALMA frequencies, we scaled the \cite{harper_2001} semi-empirical model for the M2~Iab RSG Betelgeuse (without the silicate dust) to the angular diameter of VY CMa. The main source of opacity in this model is the H$^{-}$ and H free-free opacity from electrons produced by photoionized metals. We then assumed the same ionization fraction and scaled the particle densities so that good agreement was obtained with the Very Large Array centimetre observations of VY CMa from \cite{lipscy_2005}. The stellar contribution from this weakly ionized atmosphere is 36\,mJy at 321\,GHz and 124\,mJy at 658\,GHz.

\end{appendix}
\end{document}